\documentclass[pre,twocolumn,superscriptaddress,showkeys,floatfix]{revtex4}
\usepackage{dcolumn}
\usepackage{graphicx}
\usepackage{tablefootnote}
\usepackage{amssymb}
\usepackage{amsmath}
\DeclareMathOperator{\sgn}{sgn}
\begin{document}

\date{\today}

\author{Mathias Winkler}
\email{mathias.winkler@ntnu.no}
\author{Magnus Aa. Gjennestad}
\email{magnus.aa.gjennestad@ntnu.no}
\affiliation{PoreLab, Department of Physics, Norwegian University of Science and Technology (NTNU), NO-7491 Trondheim, Norway}
\author{Dick Bedeaux}
\email{dick.bedeaux@ntnu.no}
\author{Signe Kjelstrup}
\email{signe.kjelstrup@ntnu.no}
\affiliation{PoreLab, Department of Chemistry, Norwegian University of Science and Technology (NTNU), NO-7491 Trondheim, Norway}
\author{Raffaela Cabriolu}
\email{raffaela.cabriolu@dsf.unica.it}
\affiliation{Department of Physics, Universita' degli studi di Cagliari, Complesso Universitario di Monserrato, 09042 Monserrato (CA), Italy}
\author{Alex Hansen}
\email{alex.hansen@ntnu.no}
\affiliation{PoreLab, Department of Physics, Norwegian University of Science and Technology (NTNU), NO-7491 Trondheim, Norway}

\title{Onsager-Symmetry Obeyed in Athermal Mesoscopic Systems: Two-Phase Flow in Porous Media}

\begin{abstract}
We compute the fluid flow time-correlation functions of incompressible, immiscible two-phase flow in porous media using a 2D network model.
 Given a properly chosen representative elementary volume, the flow rate distributions are Gaussian and the integrals of time correlation 
functions of the flows are found to converge to a finite value. The integrated cross-correlations become symmetric, obeying Onsager's reciprocal relations.
 These findings support the proposal of a non-equilibrium thermodynamic description for two-phase flow in porous media. 
\end{abstract}

\maketitle

\section{Introduction}

Athermal fluctuations occur in a number of phenomena in nature, important
to biology, chemistry and physics\cite{Ben-Isaac_11_238103,Gnoli_13_120601,Bi_15_63}. Currently, an active effort is taking place
to better understand the statistical physics of such systems and its use
is realized for a growing number of research areas \cite{Bi_15_63,Kanazawa_15_090601,Ben-Isaac_11_238103,Clewett_16_28726,Weber_12_7338,Dabelow_19_021009,Kumar_14_4688}. 
A particular example is granular materials, which constituents are macroscopic. In the absence of an external driving force the material will stay in its current configuration, sharing some properties with non-ergodic systems \cite{Bi_15_63}. However, when a granular material is exposed to an external force, a great number of states may be visited resulting in solid- or fluid- like behavior as a response to that force.

One area that seems to have not been analyzed in
those terms, are 
flows driven through porous media. 
Such flows are important for numerous geological and technical processes, say in oil production, CO$_2$ sequestration, water transport in aquifers, or heterogeneous catalysis. An important class of porous media flows is the simultaneous flow of two immiscible fluids. In such a system, clusters of the two fluid phases, travelling through the porous media, are constantly forced to split and recombine. Thus, the fluid configuration in the pore space changes, leading to fluctuations in the flow rate of each phase (fractional flow rate), as well as in the total flow rate. These fluctuations are of a mechanical nature, different, but analogous to thermal fluctuations on the molecular level. The fluctuations appear on a mesoscopic scale much larger than the molecular scale of statistical thermodynamics, yet the mesoscopic scale which is defined by the pore sizes of the medium is very small compared to the overall system. In the most extreme cases the pores can be in the nanosize regime, while the system of interest, for instance in chalk oil reservoirs \cite{Tang_19_827}, has geological dimensions. 

Our long-term aim is to find a non-equilibrium thermodynamic description for such flow systems. The art is then to define a suitable representative elementary volume (REV), where the essential assumption of local equilibrium, as expressed by the ergodic hypothesis, and microscopic reversibility holds. The statistical foundation of the theory was spelled out a long time ago \cite{deGroot1984}. Experimental \cite{Erpelding_13_053002} and computational \cite{Savani_17_869,Savani_17_023116} evidence, exist now, that the ergodic hypothesis can be expected to hold for immiscible two-phase flow in two-dimensional porous media of a minimum number of links. 

Here, the aim is to move one step forward, and examine the idea of time-reversal invariance or  microscopic reversibility of fluctuations \cite{Kjelstrup2008,deGroot1984}.  
Thus, our interest is the time-correlation functions of the flows. On the molecular scale, thermal fluctuations have correlation functions that are connected to transport coefficients. This is formulated in the Green--Kubo relations, which are frequently employed in molecular dynamic simulations.  
The method goes back to Onsager's regression hypothesis~\cite{Onsager_31_405,Onsager_31_2265}, which says that the decay of molecular fluctuations are governed by the same laws as the relaxation of macroscopic non-equilibrium disturbances. For the Onsager reciprocal relations \cite{Onsager_31_405,Onsager_31_2265} to apply, microscopic reversibility must hold.
The idea of the present work is to apply Green--Kubo-like relations to the fluctuations in a REV of a network model. The Green--Kubo relations for the molecular level apply to global equilibrium. In the present approach we will use similar expressions, but for fluctuations on the mesoscale, thereby extending or expanding
the Green--Kubo-scheme.  We shall see that the system models a time reversal invariant process and that the integrated flux correlations satisfy Onsager symmetry. A similar approach to fluctuations in hydrodynamic dispersion was taken by Flekkøy et al.~\cite{Flekkoy_17_022136}.

In the present study, we model the immiscible two-phase flow through a porous media 
using a hexagonal lattice, where the links represent pore throats and have a distribution in link radii. The steady flow of two incompressible 
and immiscible fluids is driven by a constant pressure difference across the network, leading to a steady state with fluctuating fluid flow.   
The flow properties fluctuate around well-defined averages and system is in a non-equilibrium steady-state on the network level of description.
A correlation of the two flows is thus unavoidable. But what is the nature of this correlation? The answer will have an impact on how we may proceed with a theoretical description of the flows.  

If one considers a steady state of the immiscible
two-phase flow, as we will in our model, we shall see that the fluctuations become Gaussian around a steady mean.

Hence, the concept of the REV at steady state is highly relevant and important for how we build a theory that can help us understand transport in porous media.

\begin{figure}[htb]
\includegraphics[width=0.5\textwidth]{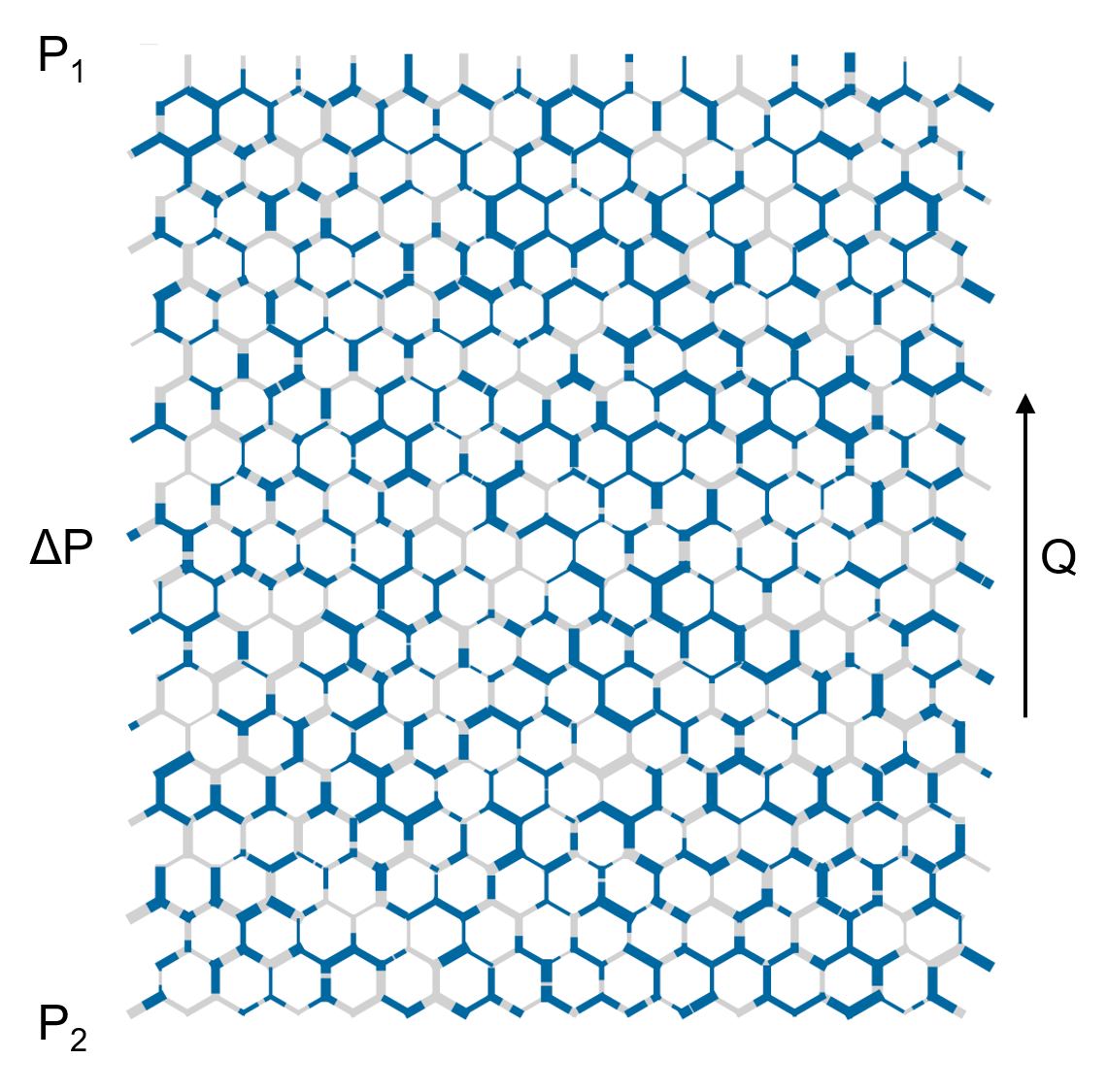}
\caption{\label{fig:model}
Illustration of the network model. The network is occupied by two immiscible fluids (blue and gray colors).
The equally long links (pores) have hourglass shaped structure and a random distribution of diameters.}
\end{figure}

\section{Model}

The transport of the two immiscible fluids through a two-dimensional porous medium is represented by a 
dynamical pore network model \cite{Aker_98_163,Sinha_19_190712842S}. This model has been in development over two decades and
has a record of explaining experimental and theoretical results in steady and transient two-phase flow in porous media \cite{Aker_98_163,Savani_17_023116,Savani_17_869,Sinha_19_190712842S,Sinha_17_77,Erpelding_13_053002,Gjennestad_18_56,Zhao_19_01619,Santanu_19_65}. In this model the two fluids are separated by interfaces and
are flowing in a network of links which are connected at nodes. The network has a honeycomb structure as illustrated in Figure~\ref{fig:model}, with equally long links and a distribution of link radii. The radii are drawn from a uniform distribution in the interval
0.1$L$ to 0.4$L$, where $L$ is the length of the links. The flow rate $q_{ij}$ inside a link connecting nodes $j$ and $i$ is given by:
\begin{equation}
\label{eq:linkflow}
q_{ij} = -g_{ij}[p_j-p_i-c_{ij}(\mathbf{z_{ij}})].
\end{equation}
Here $p_j$ and $p_i$ are the pressures at nodes $j$ and $i$, $c_{ij}$ is the capillary pressure, and $g_{ij}$ is the conductivity of the link. The links have an hourglass shape, thus
the capillary pressure is a function of the interface positions, $\mathbf{z_{ij}}$:
\begin{equation}
    \label{eq:interface}
    c_{ij}(\mathbf{z_{ij}}) = \frac{2\gamma}{r_{ij}} \sum_{z\in\mathbf{z_{ij}}} (\pm 1)\left\{1 - \cos (2\pi\chi(z)) \right\}.
\end{equation}
Here $\gamma$ is the surface tension, $r_{ij}$ is the radius of link ij, and
\begin{equation}
    \chi(z) = \left\{\begin{array} {ll} {0,} & {\text{if} \ z < \beta r_{ij}}, \\ 
    {\frac{z-\beta r_{ij}}{L - 2\beta r_{ij}},} & { \text{if}\  \beta r_{ij} < z < L - \beta r_{ij}, } \\ 
    {1,} & {\text{if}\  z > L - \beta r_{ij}}. \end{array} \right.
\end{equation}
The effect of the $\chi$-function is to introduce zones of length $\beta r_{ij}$ at each end of
the links where the pressure discontinuity of any interface is zero.
The conductivity of the link, $g_{ij}$, contains a geometrical factor and the effective viscosity of the link:
\begin{equation}
    g_{ij} = \frac{\pi r_{ij}^4}{8 L \mu(S_{w,ij})}.
\end{equation}
Here, $r_{ij}$ is the radius of the link and the viscosity is defined as
\begin{equation}
   \label{eq:linksat}
    \mu(S_{w,ij}) = S_{w,ij}\mu_w + (1-S_{w,ij})\mu_n, 
\end{equation}
with $S_{w,ij}$ being the saturation (i.e. the volume fraction) of the wetting phase in link $ij$.
Simulations were carried out, applying a constant global pressure drop $\Delta{P}$ across the network. Periodic boundary conditions are used in all directions. The local pressures $p_i$ are determined by solving the Kirchhoff equations. Further details
 of the model and solution methods can be found in Refs.~\citenum{Sinha_19_190712842S,Gjennestad_18_56}.
For each link the flow rate $q_{ij}$ is calculated using equation \ref{eq:linkflow} and the positions of the interfaces are advanced with an appropriately small, constant time step of 10$^{-5}$ s.
A constant time step is used to facilitate a convenient calculation of the time-correlation functions. The simulations were started with a random distribution of the two liquid phases, and were
propagated at least 300000 time steps to allow for the system to reach steady state. 
Statistics for the time correlation functions were collected for 9.7 million time steps. The length of a link in the network was set to 1mm.
We report results for each set of parameters as averages of at least 30 runs using different starting configurations of the two fluids. Volume flow rates and velocities refer to network averages.
The properties of the steady state flow are determined by the pressure drop across the system, $\Delta{P}$, the total wetting saturation of the network, the surface tension, $\gamma$, and the viscosity of the two fluids. We have ensured that the same steady state flows averages are obtained from different initial distributions of the two fluids including an initial configurations where the two phases are completely separated (i.e. each phase is in a single connected cluster). Furthermore, it has been tested that a different link radii configuration, drawn from the same uniform distribution does not change the steady flow averages nor the appearance of the time-correlation functions computed in this study.

We investigated time correlation functions and their long-time convergence for two choices of the parameter set $\mu_w$, $\mu_n$ and $\gamma$, which are viscosities of the wetting and the nonwetting phase and the surface tension, respectively. 

Case (A) had viscosities  $\mu_w=\mu_n=10^{-3}$ Pa$\cdot$s, different choices for the pressure gradient (100-200 kPa/m) and the surface tension $\gamma$ = 3 - 30 mN/m. At steady state conditions, the capillary number was $\mathrm{Ca}\approx$ 10$^{-3}$ - 10$^{-2}$. Here, the capillary number is defined as $\mathrm{Ca} = v \mu_{n}/\gamma$, with $v$ being the seepage velocity. The case was chosen to represent flow where the two fluids are interchangeable with respect to their viscous dissipation.   

Case (B) had $\mu_w=5\mu_n$ ($\mu_w=10^{-3}$ Pa$\cdot$s) and $\gamma$ = 0. This case is typical at high flow rates where the contribution from $\gamma$ essentially can be neglected, and the capillary number goes to infinity. It can be considered as a limiting case, chosen to elucidate the behavior when viscous forces dominate and the surface tension is negligible. Data was collected for wetting phase saturation S$_w$=0.25, 0.5 and 0.75. Here, S$_w$ is the volume fraction of the wetting phase of the total pore volume in the network. 


\section{Results and Discussion}

We report first that the fluctuations in flow velocities are Gaussian
when a suitable representative volume (REV) is chosen.  
We proceed to give the structure of the time correlation functions for the REV. The results for what we will call from now the Green-Kubo coefficients for the network follow from this. 

\subsection{Fluctuations}

In case (A) the resistance is determined
by the positions of the interfaces in the links only, as the two phases have the same viscosity.  In case (B), the resistance
to flow in link $i$ is inversely proportional to the effective viscosity. The positions of the interfaces are then irrelevant as there is no surface tension and hence, no capillary pressure.

\begin{figure}[htb]
\includegraphics[width=0.5\textwidth]{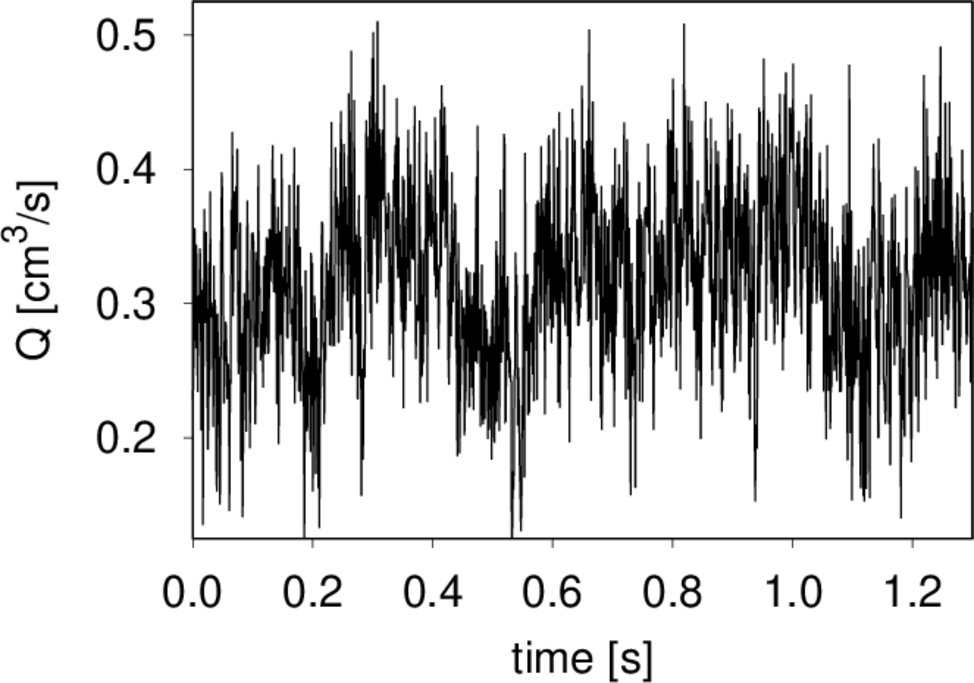}
\caption{\label{fig:fluctuations} Fluctuations of the volumetric flow Q for case A (see text).
The pressure gradient was 100 kPa/m. 
During the time span shown, 1.3 seconds, a volume twice the total volume of the network has passed. }
\end{figure}

\begin{figure}[htb]
\includegraphics{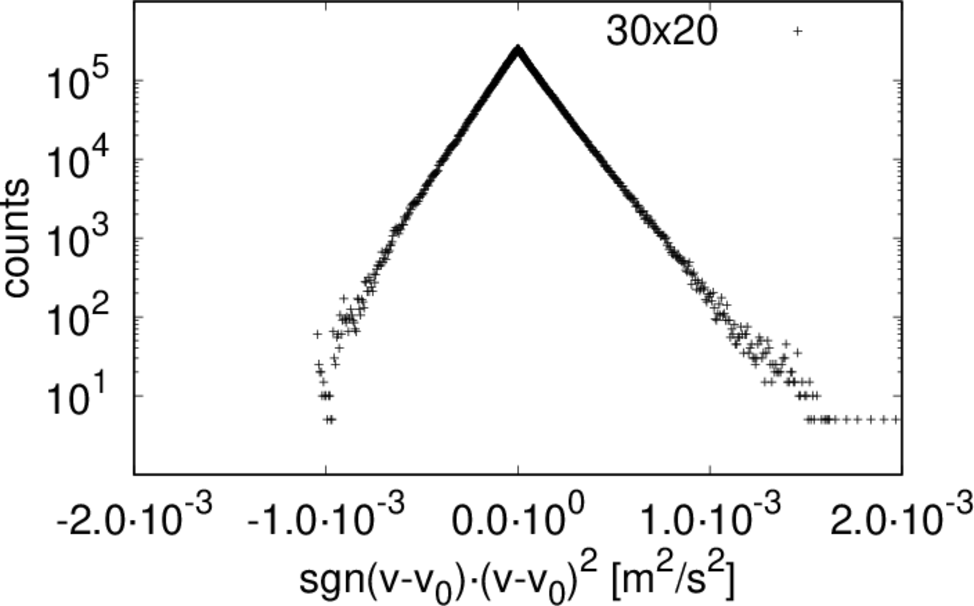}
\caption{\label{fig:gauss_log} Distribution of network-averaged instantaneous total (wetting and nonwetting) fluid velocity for network size 30x20 (case A). $v_0$ is the  total average velocity and sgn is the sign function.}
\end{figure}

\begin{figure}[htb]
\includegraphics{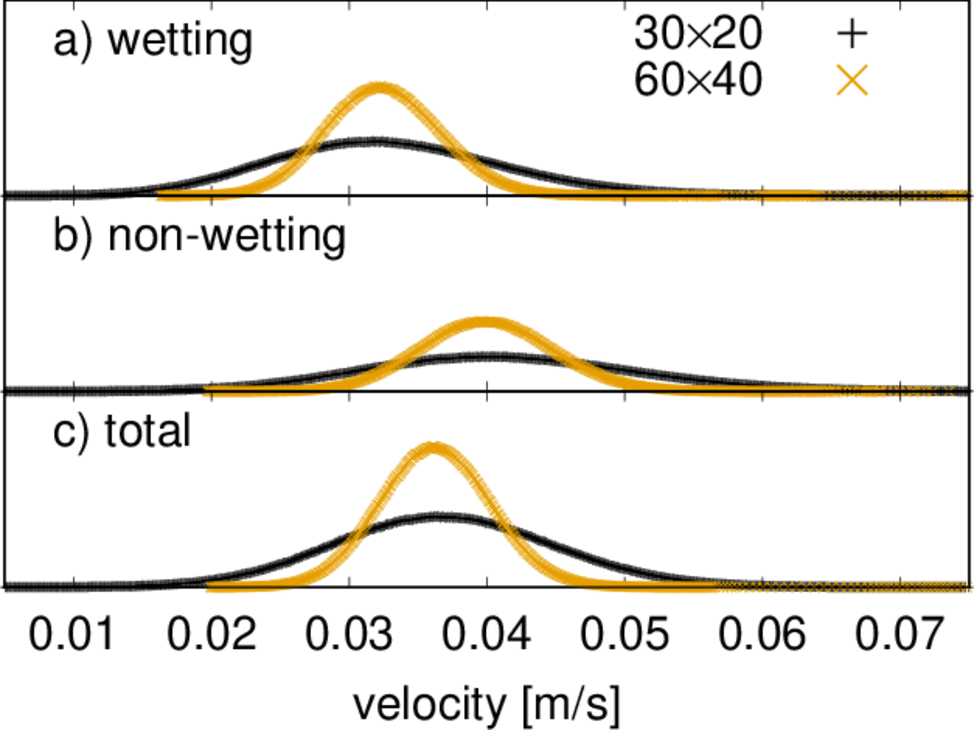}
\caption{\label{fig:vel_dist} Distributions of network-averaged instantaneous fluid velocities for two network sizes (30$\times$20, 60$\times$40). }
\end{figure}

A typical example of fluctuations in the total volume flow $Q$ for case (A), with a pressure gradient ($\Delta{P}/\Delta{x}$ of 100 kPa/m and surface tension 30 mN/m, is shown in Figure~\ref{fig:fluctuations}.
By plotting the statistical frequency of the flow rate or the seepage velocity, we obtain 
a Gaussian distribution. This is shown in Figure~\ref{fig:gauss_log} where we plot the statistical frequency of the fluid velocity (counts) on a logarithmic scale vs.~$\sgn(v-v_0)(v-v_0)^2$. In such a plot a Gaussian distribution appears as a triangle, and this behavior is very well followed by the data. There is only a very small asymmetry in the distribution which is to be expected as the fluid velocity cannot be less than zero. A regular plot of the distributions is presented for the seepage velocity, and
the velocities of the wetting and non-wetting phases in Figure~\ref{fig:vel_dist}. 
The distributions are shown for two network sizes of case (A), with 30$\times$20 links and one twice the size of the former, 60$\times$40 links. 
The shown distributions are normalized with the area, and the variance of the 
larger network is 1/2 the width of the smaller network. So, in spite of the apparent noise seen in Figure~\ref{fig:fluctuations}, one obtains the distributions in Figure~\ref{fig:vel_dist}, which has its analogue in a molecular picture, basic to thermodynamic equilibrium properties. This allows us to proceed with the next step, and construct time correlation functions for the meso-level.

\begin{figure}[htb]
\includegraphics{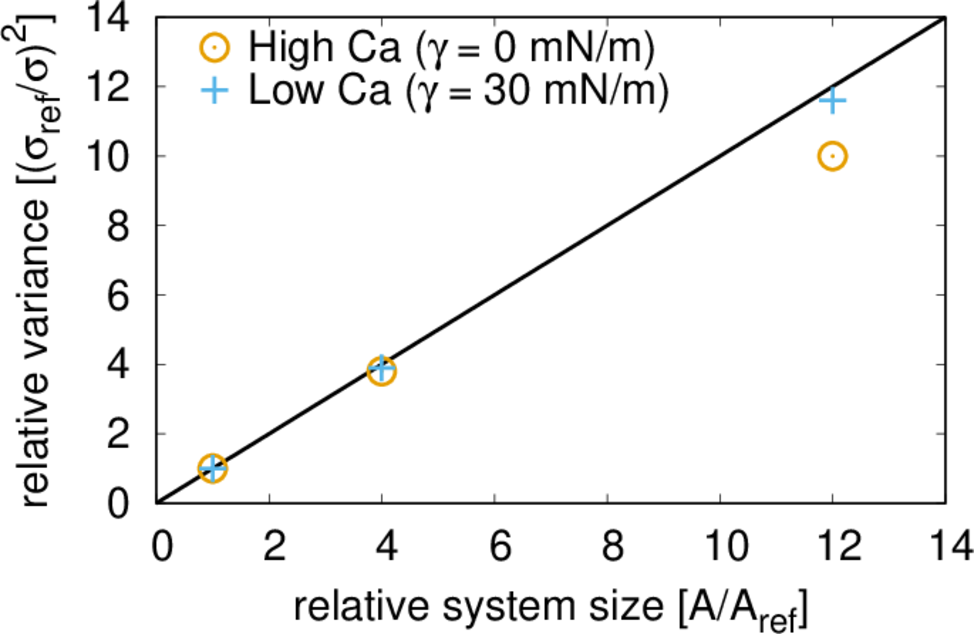}
\caption{\label{fig:sig2_size} Scaling of the variance $\sigma^2$ of the flow velocity size distributions with system size.}
\end{figure}


\subsection{Network Size and Representative Volume}

Ideally, the system size of the simulation is sufficiently large and representative of the statistical ensemble. In this the entropy and other thermodynamic properties are proportional to the system size, (i.e. they are extensive \cite{Kjelstrup_19_150}). 
With the Gaussian nature
one may expect that the inverse variance 1/$\sigma^2$ 
of the fluctuations is proportional to the system size, i.e the area A. This relation is plotted
in Figure~\ref{fig:sig2_size} for network models of dimension 30$\times$20, 60$\times$40 and 120$\times$60 links. It shows that this requirement is well met by a system with low Ca, but systems with higher Ca may be more susceptible to possible size effects. 
The size of the REV will be system dependent, see Savani et al.~\cite{Savani_17_023116}. But in the present cases (A) and (B), a REV can be defined, for a range of Ca, different for the different cases. The results for the REV complies with the meso-level analog we are seeking. 

\subsection{Time Correlation Functions}

With a well-defined REV, and with Gaussian fluctuations established, we can proceed to define the time correlation functions $C_{RS}$ for the fluctuating quantities $R$ and $S$ at the meso-level:
\begin{equation}
\label{equ:TCF}
C_{RS}(\tau) = \langle\delta R(0)\delta S(\tau)\rangle = \langle R(0)S(\tau) \rangle - \langle R \rangle \langle S \rangle,
\end{equation}
\begin{figure}[htb]
\includegraphics{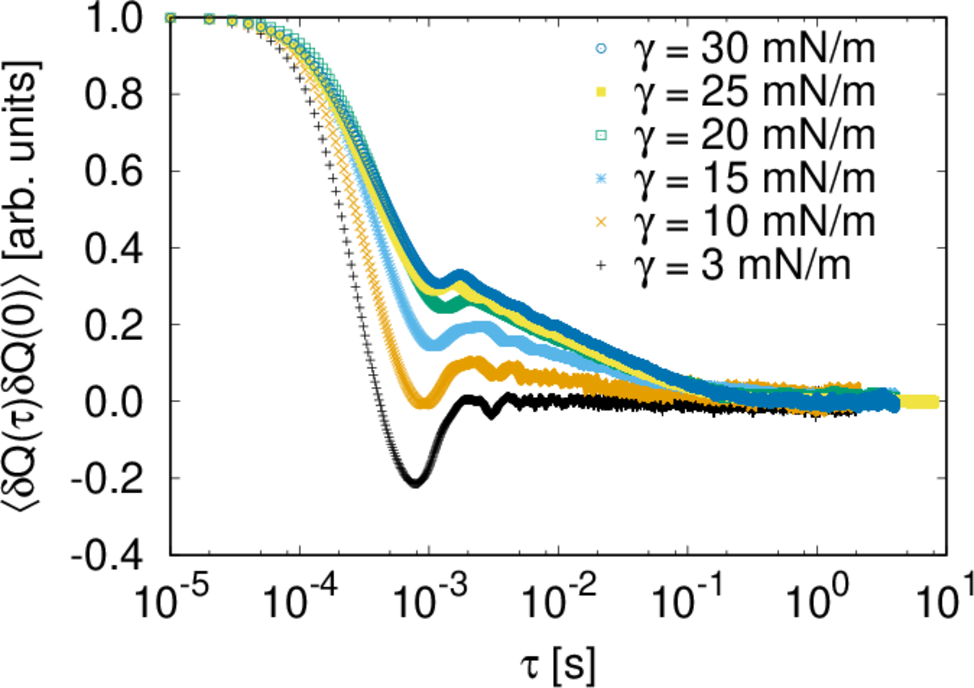}
\caption{\label{fig:QQ_sigs} Dependence of (scaled) time correlation function $\langle Q(\tau)Q(0) \rangle$ on the surface
tension $\gamma$ (case (A) with $\Delta{P}/\Delta{x}$ = 100 kPa/m and network size of 30$\times$20 links).}
\end{figure}
\begin{figure}[htb]
\includegraphics{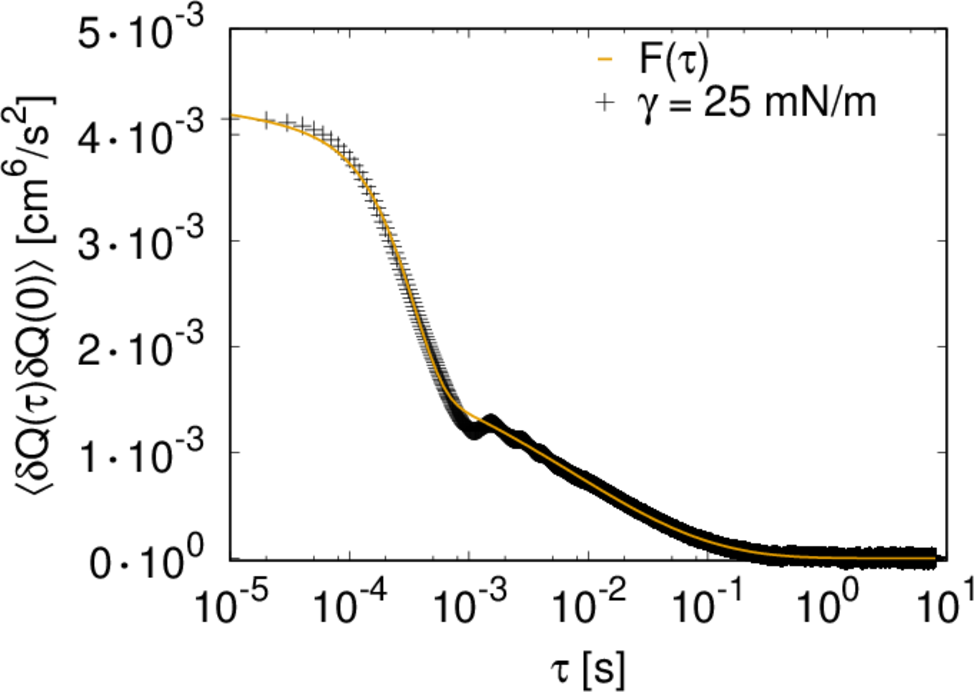}
\caption{\label{fig:fit} Fit of function F(t) (equation~\ref{equ:fit}) to the
autocorrelation function of the total volume flow for case (A) with $\gamma$ = 25 mN/m 
($\Delta{P}/\Delta{x}$ = 100 kPa/m and 30$\times$20 links).}
\end{figure}
where the brackets $\langle \cdots \rangle$ indicate ensemble averages. The fluctuation from the mean, $\delta R$, is defined as
\begin{equation}
    \delta R(t) = R(t) - \langle R \rangle,
\end{equation}
and
\begin{equation}
    \langle \delta R \rangle = 0.
\end{equation}
Figure~\ref{fig:QQ_sigs} shows the time correlation functions of the total flow rate $Q$ for different choices of the surface tension. After a rapid decay on a short timescale (below 10$^{-3}$ s) there is a slower, logarithmic
decay which is more pronounced for larger values of $\gamma$ (between 1 ms and 100 ms). These two regimes
are followed by a slow long time decay. The rapid decay appears on the time scale that correspond to
the time necessary to evolve the flow by one average link volume. As shown in Figure~\ref{fig:QQ_sigs}, 
the decay is somewhat faster for smaller surface tensions as the total flow velocity is higher. 

The regime of the logarithmic decay is within the time of evolving the flow by the total volume of the network,
and is more pronounced for higher surface tensions and thus higher capillary pressures in the pores. Hence, the decay corresponds to parts of the flow that is slow moving or frustrated. These are the regimes of interest here. They contain the relative movements of the two flows in terms of their mutual displacement. 

It is interesting to note some similarities with time correlation functions of glass\cite{Reichman_05_P05013} or yield-stress fluids\cite{Levashov_17_184502}.
In these cases, the autocorrelation functions, like the self-scattering function, can be fitted to a  function of the form: 
\begin{equation}
\label{equ:fit}
F(t) = a \exp[-(t/\tau_1)^\alpha] + b \exp[-(t/\tau_2)^\beta].
\end{equation}
We attempted a fit of $F(t)$ to the autocorrelation functions of the total flow, see Figure~\ref{fig:fit}. Satisfying fits
could be obtained with the exception that the local minimum and maximum at around 10$^{-3}$ s and in some cases the flat top (at times $<$ 10$^{-4}$ s) are not well described. Fit parameters for the
different choices of $\gamma$ are summarized in Table~\ref{tbl:fit}.


\begin{table}
\caption{\label{tbl:fit} Fitting parameter for $F(t)$ (see equation~\ref{equ:fit}) to the autocorrelation functions
of the total flow for different values of $\gamma$. The units of the fitting parameters are [cm$^6$/s$^2$]$\cdot$10$^{-2}$ ($a$ and $b$),
and [ms] ($\tau_1$, $\tau_2$).}
\begin{tabular}{lcccccc}
\hline
$\gamma$ [mN/m] & $a$  & $b$ & $\tau_1$ & $\tau_2$ & $\alpha$ & $\beta$ \\
\hline
30  &	0.28&	0.17&	0.39&	22&	1.40&	0.52 \\
25  &  0.21&	0.23&	0.34&	6.6&	1.75&	0.36 \\
20  &  0.13&	0.50&	0.49&	0.059&	2.69&	0.16 \\
15  & 	0.15&	0.22&	0.39& 	0.096&	2.28&	0.16 \\
10  &  0.11&	0.043&	0.31&	0.26&	2.09&	0.15 \\
\hline
\end{tabular}
\end{table}

\subsection{Convergence and Symmetry}

\begin{figure}[htb]
\includegraphics{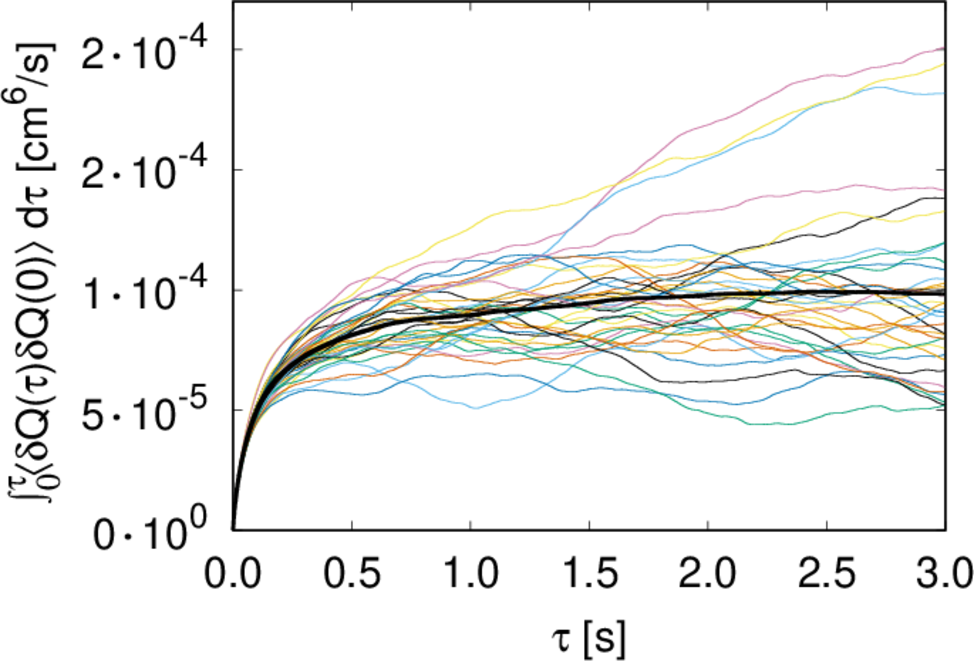}
\caption{\label{fig:Convergence} Convergence of the integrated time correlation function (case (A), $\gamma$ = 30 mN/m and 30$\times$20 links). The coloured lines represent
individual trajectories, the thicker black line is the average of all trajectories.}
\end{figure}

The Green--Kubo method employs integrals of suitable time-correlation functions C$_{RS}$ (as defined in equation~\ref{equ:TCF}) to compute coefficients, $L_{RS}$:
\begin{equation}
L_{RS} = \int_0^{\infty} C_{RS}(\tau) d\tau.
\end{equation}
The convergence of the integral over the time correlation function of the total flow  is shown in Figure~\ref{fig:Convergence}. As in molecular dynamics, where the Green--Kubo method is normally used, 
the convergence is slow and statistics has to be collected over long time-scales and/or multiple trajectories
to achieve convergence of the integral when $\tau$ is approaching infinity.
\begin{figure*}[htb]
\includegraphics[width=\textwidth]{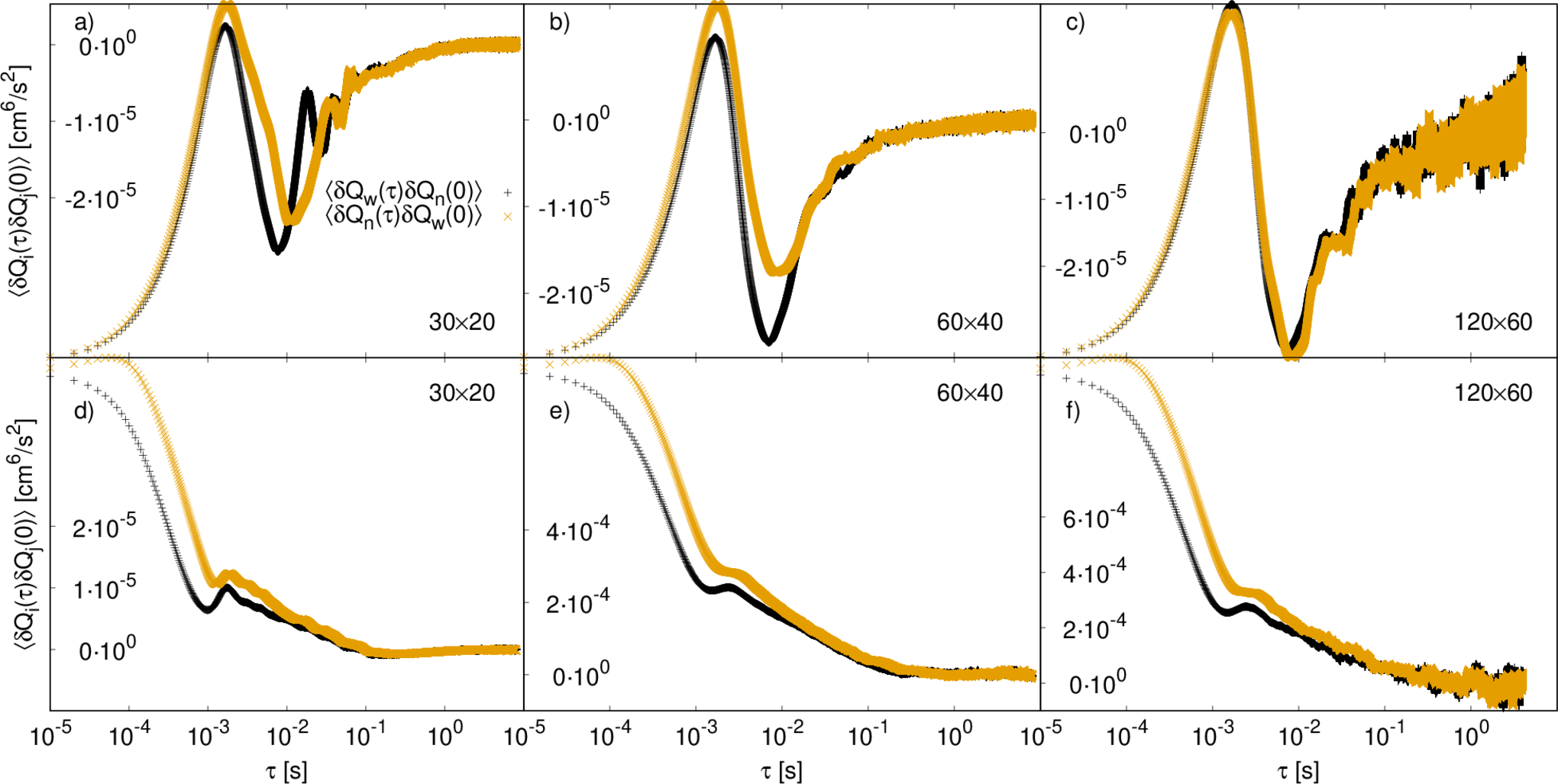}
\caption{\label{fig:Cross_Size} Cross-correlation functions of wetting (Q$_w$) and nonwetting (Q$_n$) flows. The upper 
panels a)-c) show the cross-correlations for case (B) with zero surface tension and increasing network size from a) to c).
The lower panels d)-f) show the cross-correlations for case (A) with $\gamma$ = 30 mN/m. The network size is increasing
from left to right 30$\times$20, 60$\times$40 and 120$\times$60 links. }
\end{figure*}

We computed the integrals for the autocorrelation and crosscorrelation functions of the wetting and nonwetting phases,
\begin{equation}
\label{equ:lambda}
\Lambda_{ij} = \int_0^{\infty} [\langle Q_i(\tau)Q_j(0) \rangle - \langle Q_i \rangle \langle Q_j \rangle]  d\tau,
\end{equation}
with the indexes $i,j = n,w$ referring to the nonwetting and wetting phase, respectively. The results are listed in Table~\ref{tbl:LambdalowCa} for case (A), where the surface tension is 30 mN/m and the fluids have the same viscosity, for three different choices of $\Delta{P}$, the pressure difference across the network. Table~\ref{tbl:LambdahighCa} summarizes results for infinite capillary numbers (case B), where the surface tension is zero but the fluids have different viscosities, for different choices of the saturation. 

In all cases (A) and (B) we find that 
the integral cross correlations obey the 
Onsager reciprocal relation, $\Lambda_{ij} = \Lambda_{ji}$, within the statistical error in the simulations. This result is new for a meso-level description, like the one used here, and is encouraging for the overall aim; to create a non-equilibrium thermodynamic description on for the macroscopic level. The finding applies to a well defined REV, for which we have a Gaussian distribution of fluctuations, analogous to the corresponding distribution on the molecular level.  

It is interesting that the cross coefficients are all negative. This makes sense for network flow where one component cannot advance faster (on average) than the mean flow, unless the other component advances slower (on average). 

The $\Lambda_{ij}$s for case (B), where the surface tension is zero, show an extreme limit property, because the
crosscorrelation functions obey $\Lambda_{ww}\Lambda_{nn} - \Lambda_{wn}\Lambda_{nw} \approx 0$. A singular matrix of coefficients
is the essence of complete coupling of the two fluids' flows; they are linearly dependent. For all choices of saturation $\Lambda_{ww} = \zeta^2 \Lambda_{nw}$, where $\zeta$ is a constant \cite{Kjelstrup2008}. 
On the other hand, for case (A), where the surface tension differs from zero, a deviation of this dependency is observed. The linear dependence of the fluxes in case (B), can thus be associated with lack of capillary forces. This can be understood in the following way: in case (B) the variation of mobility in a given link is a function of the saturation in the link only. However, if the mobility in one link is increased it has to decrease elsewhere. 
In contrast, for case (A) the variations in link mobility depend on the interface position and a change in the link mobility can take place without affecting the mobilities of other links. 

The value of $\zeta$ for case (B) can be deduced by looking at the coefficients in Table IV. Within the accuracy, we find $\zeta^2 \approx 20$ or $\zeta = 4.5 \pm 0.5$ for all $S_w$. The value is close to the ratio of fluid viscosities, which will describe the dissipation.

All coefficients in Table IV show a dependence on the saturation, decreasing in value as the saturation is increasing. Here, the wetting fluid is the more viscous fluid and the increase in saturation reduces the effective permeability. The coefficients show also a dependence on the pressure drops across the network (see Table II), increasing with higher pressure drops. This is consistent with a higher effective permeability at higher pressures. In fact the dependence of the volumetric flow is non-linear as can be seen in Table III. A non-linear dependence of a flow rate on the pressure difference is a well-known phenomena in immiscible two-phase flow\cite{Longeron_80_391}.   

To investigate the origin of the Onsager symmetry in more detail we examined the cross correlations in Figure~\ref{fig:Cross_Size}. For molecular systems, one can find transport coefficients using the Green--Kubo method, see \cite{Liu_13_1169} and Onsager reciprocal relations apply given time reversal invariance. We have found that time reversal invariance applies also on the mesoscopic level, as formulated by: 
$C_{AB}(\tau) = C_{BA}(\tau)$. 
This equality is pictured in Figure~\ref{fig:Cross_Size}. There is agreement, except for very small times where the contribution to the integral 
is negligible. Moreover for case (B), the deviation from symmetry at small times is attributable to the finite system size. It is vanishing for larger network sizes. This is shown in the upper three panels of Figure~\ref{fig:Cross_Size}. 

\begin{table}
\caption{\label{tbl:LambdalowCa} Integrated time correlation functions for case (A) and three different settings of the
pressure drop across the network. $\Lambda{_{ij}}$ are obtained from equation~\ref{equ:lambda}.
The uncertainties for $\Lambda_{i,j}$ are estimated to be less than 21\%.}
\begin{tabular}{lccc}
\hline
$\Delta{P}/\Delta{x}$ [kPa/m] & 100 & 150 & 200 \\
\hline
$\Lambda{_{ww}}$ [cm$^6$/s$]\cdot$10$^{-4}$  & 0.46   & 0.72 & 1.59 \\             
$\Lambda{_{nn}}$ [cm$^6$/s$]\cdot$10$^{-4}$  & 0.69   & 1.51 & 1.88 \\             
$\Lambda{_{wn}}$ [cm$^6$/s$]\cdot$10$^{-4}$  & -0.11 & -0.35 & -0.73 \\           
$\Lambda{_{nw}}$ [cm$^6$/s$]\cdot$10$^{-4}$  & -0.10 & -0.32 & -0.70 \\           
\hline
\end{tabular}
\end{table}

\begin{table}
\caption{\label{tbl:flowlowCa} Volumetric flow rates Q and fluid velocities v for case (A) and three different settings of the
pressure drop across the network. }
\begin{tabular}{lccc}
\hline
$\Delta{P}/\Delta{x}$ [kPa/m] & 100 & 150 & 200 \\
\hline
Q [cm$^3$/s]  & 0.308 & 0.869 & 1.565 \\
Q$_w$ [cm$^3$/s]  & 0.139 & 0.401 & 0.723 \\
Q$_n$ [cm$^3$/s]  & 0.169 & 0.467 & 0.841 \\
v [m/s] & 0.037  & 0.104  & 0.188 \\
v$_w$ [m/s] & 0.033  & 0.096 & 0.174 \\
v$_n$ [m/s] & 0.041 & 0.113 & 0.203 \\
v$_n$-v$_w$ [m/s] & 0.007 & 0.017 & 0.028 \\
\hline
\end{tabular}
\end{table}

\begin{table}
\caption{\label{tbl:LambdahighCa} Integrated time correlation functions for case (B) and three different choices for the
saturation. $\Lambda{_{ij}}$ are obtained from equation~\ref{equ:lambda}. Uncertainties for $\Lambda_{i,j}$ are estimated to be less than 21\%.  }
\begin{tabular}{lccc}
\hline
$S_w$ & 0.25 & 0.5 & 0.75 \\
\hline
$\Lambda{_{ww}}$  [cm$^6$/s$]\cdot$10$^{-4}$  & 0.011  & 0.006  & 0.001\\
$\Lambda{_{nn}}$  [cm$^6$/s$]\cdot$10$^{-4}$  & 0.212  & 0.113 & 0.020 \\
$\Lambda{_{wn}}$  [cm$^6$/s$]\cdot$10$^{-4}$  & -0.046 & -0.024 & -0.005  \\
$\Lambda{_{nw}}$  [cm$^6$/s$]\cdot$10$^{-4}$  & -0.048 & -0.027 & -0.005 \\
\hline
\end{tabular}
\end{table}
\begin{table}
\caption{\label{tbl:flowhighCa}  Volumetric flow rates Q and fluid velocities v for case (B) and three different choices for the
saturation. }
\begin{tabular}{lccc}
\hline
$S_w$ & 0.25 & 0.5 & 0.75 \\
\hline
Q [m$^3$/s]  & 0.763 & 0.503 & 0.359 \\
Q$_w$ [cm$^3$/s]  & 0.137 & 0.208 & 0.248 \\
Q$_n$ [cm$^3$/s]  & 0.626 & 0.295 & 0.110 \\
v [m/s] & 0.92   & 0.61  & 0.44  \\
v$_w$ [m/s] & 0.066  & 0.050 & 0.040 \\
v$_n$ [m/s] & 0.10 & 0.072 & 0.055 \\
v$_n$-v$_w$ [m/s] & 0.034 & 0.022 & 0.015 \\
\hline
\end{tabular}
\end{table}

\section{Conclusions}

Our investigation of time correlation functions has revealed interesting parallels between the time correlation functions of two immiscible fluids in a porous media, those observed for glass and stress- yield fluids, and those for molecular fluctuations. A network with incompressible fluids has been used as model for the porous medium, but the findings should not be restricted to this. We have been able for the first time to find Onsager symmetry in athermal fluctuations on the meso-level. The symmetry of the coefficients implies time reversal invariance or microscopic reversibility of fluctuations also on the meso-level, in agreement with recent experimental \cite{Erpelding_13_053002} and computational evidence \cite{Savani_17_023116,Savani_17_869}. Time reversal invariance is here understood as $C_{AB}(\tau) = C_{BA}(\tau)$, holding for all time scales except very short times.

We found that the structure of the time correlation functions depends on the surface tension.  
Integrals over auto and crosscorrelation functions of a REV, were found to converge and the integrals 
of the crosscorrelation functions essentially obeyed Onsager's symmetry. The  coefficients obtained in this manner may have a relation to porous media permeabilities. Further research of time correlation functions to compute transport properties of immiscible-two phase flow is therefore encouraged.


\begin{acknowledgments}
The authors would like to thank Prof.~Daan Frenkel for helpful discussions of this work. This work was supported by the Research Council of Norway through its Centres of Excellence funding scheme, project number 262644. MW is grateful for a postdoc scholarship from the Department of Physics, Norwegian University of Science and Technology, NTNU, Trondheim. 
\end{acknowledgments}



\bibliography{Literature}

\end{document}